# Seneca: Taint-Based Call Graph Construction for Java Object Deserialization

JOANNA C. S. SANTOS, University of Notre Dame, USA
MEHDI MIRAKHORLI, University of Hawaii at Manoa, USA
ALI SHOKRI, Virginia Tech, USA

Object serialization and deserialization are widely used for storing and preserving objects in files, memory, or database as well as for transporting them across machines, enabling remote interaction among processes and many more. This mechanism relies on reflection, a dynamic language that introduces serious challenges for static analyses. Current state-of-the-art call graph construction algorithms do not fully support object serialization/deserialization, i.e., they are unable to uncover the callback methods that are invoked when objects are serialized and deserialized. Since call graphs are a core data structure for multiple types of analysis (e.g., vulnerability detection), an appropriate analysis cannot be performed since the call graph does not capture hidden (vulnerable) paths that occur via callback methods. In this paper, we present Seneca, an approach for handling serialization with improved soundness in the context of call graph construction. Our approach relies on taint analysis and API modeling to construct sound call graphs. We evaluated our approach with respect to soundness, precision, performance, and usefulness in detecting untrusted object deserialization vulnerabilities. Our results show that Seneca can create sound call graphs with respect to serialization features. The resulting call graphs do not incur significant runtime overhead and were shown to be useful for performing identification of vulnerable paths caused by untrusted object deserialization.

CCS Concepts: • **Software and its engineering** → **Automated static analysis**; *Software verification and validation.*

Additional Key Words and Phrases: object serialization, untrusted object deserialization, taint analysis, call graphs

## 1 INTRODUCTION

Static program analysis is a key component of today's software analysis tools that bring automation into activities such as defect localization and/or finding (*e.g.*, [Dolby et al. 2007; Thaller et al. 2020]), vulnerability detection (*e.g.*, [Jovanovic et al. 2006; Liu Ping et al. 2011]), information flow analysis [Sridharan et al. 2011], code refactoring (*e.g.*, [Khatchadourian et al. 2019]), code navigation (*e.g.*, [Feldthaus et al. 2013]), code clone finding (*e.g.*, [Wyrich and Bogner 2019]), and optimization [Hines et al. 2005]. Such tools often perform multiple types of inter-procedural analysis, that leverage ***call graphs*** – data structures that indicate caller-callee relationships [Grove

Authors' addresses: Joanna C. S. Santos, Department of Computer Science and Engineering, University of Notre Dame, Notre Dame, IN, 46556, USA, joannacss@nd.edu; Mehdi Mirakhorli, Department of Information and Computer Sciences, University of Hawaii at Manoa, Honolulu, HI, 96822, USA, mehdi23@hawaii.edu; Ali Shokri, Department of Electrical and Computer Engineering, Virginia Tech, Blacksburg, VA, 24061, USA, Email:ashokri@vt.edu.

and Chambers 2001]. However, prior works have demonstrated that constructing a call graph for object-oriented programs is often non-trivial, expensive and/or non-feasible due to the usage of many dynamic programming language constructs. For instance, native calls, reflection, and object serialization make it challenging to statically construct a **sound** call graph [Ali et al. 2019; Kummita et al. 2021; Reif et al. 2019, 2018; Smaragdakis et al. 2015; Sridharan et al. 2013].

These programming constructs are heavily used in contemporary software systems as they enable the developers to link/load new class libraries, methods, and objects and extend the programs' functionalities [Landman et al. 2017; Reif et al. 2019]. Ignoring such constructs leads to *unsound* call graphs in which feasible runtime paths are missed, and call graphs cannot be used to infer the possible execution from the code [Reif et al. 2019, 2018; Sridharan et al. 2013]. To tackle this problem, previous works explored certain classes of language features, such as reflection features [Bodden et al. 2011; Li et al. 2014, 2019; Smaragdakis et al. 2015], native (opaque) code [Smaragdakis et al. 2015], dynamic proxies [Fourtounis et al. 2018], and programs with Remote Method Invocation (RMI) [Sharp and Rountev 2006]. However, as demonstrated by Reif et al. [Reif et al. 2019, 2018], a powerful and frequently used programming construct that has been left out from the programming analysis techniques is *serialization (and deserialization) of objects*.

Object *serialization* is the process of converting (the state of) an object into an abstract representation (*e.g.*, a byte stream or JSON, *etc.*). The reverse process of reconstructing objects from its abstract representation is called *deserialization*. This is a widely used mechanism for storing and preserving objects in files, memory, or database as well as for transporting them across machines, enabling remote interaction among processes and many more. For example, the Android API provides a `Bundle` object which can be used for inter-process communication between apps as well as Android's OS with an individual app via their serialization and deserialization [Arzt et al. 2014; Enck et al. 2014]. Moreover, object (de)serialization is also used to improve the system's performance by saving objects for later retrieval, *e.g.*, saving a trained machine learning model to be used later without the need to retrain the algorithm. Serializing an object has other advantages, such as being readable by applications in other languages. For instance, JavaScript running in a web browser can natively serialize and deserialize objects to and from JSON, therefore interact with other applications written non-JavaScript languages.

Although object serialization is widely used in many languages and commonly adopted by programmers, static analyzers do not fully cover analysis of programs with this construct yet [Reif et al. 2019, 2018]. This is particularly important considering the spike of vulnerabilities related to *untrusted object deserialization* [Muñoz and Schneider 2018; Sayar et al. 2023; Schneider and Muñoz 2016] that cannot be automatically detected because call graphs are unsound. For example, Apache's Log4j software library (versions 2.0-beta9 to 2.14.1) had an untrusted object deserialization vulnerability that allowed remote code execution. This was a critical vulnerability that affected several software systems.

As demonstrated by previous studies on the *soundness* of call graph construction approaches [Reif et al. 2019, 2018]— *guaranteeing that all possible behaviors are modeled in a call graph* — state-of-the-art techniques do not support serialization-related operations. They fall short in having nodes and edges that represent *callback* methods that are invoked during the serialization or deserialization of objects. There are multiple reasons on why it is hard to handle this language construct:

— Serialization and deserialization use several overridable callback method(s). These call back methods are invoked by the Java API using "non-trivial" reflective calls that current techniques [Landman et al. 2017] for taming reflection do not address. Therefore, the resulting call graph underapproximates the program's behavior; they miss potential program paths through these call back methods.

— The invoked callbacks during deserialization methods depend on the received object, which is coming from an external stream. The values and internal field types are only known at runtime when the object deserialization occurs.
— The external stream may include objects whose types are not observed statically, *i.e.*, they are available in the classpath (imported libraries, or Java built-in API) but were never actually used (instantiated) in the application scope. A typical static analysis would consider these types as unused.

Therefore, existing techniques on addressing reflections had failed to address call-graph generation with the presence of object serialization/de-serialization [Reif et al. 2019]. As such, potential program flows are disregarded in existing call graph construction algorithms. Since the call graph is a core data structure in performing many inter-procedural code analyses, the underlying client would suffer with the *unsoundness*. In use-cases such as detection of untrusted deserialization vulnerabilities, an appropriate analysis cannot be performed since the call graph does not capture hidden (vulnerable) paths that occur via callback methods. There are two algorithms that (partially) handle serialization constructs (*i.e.*, CHA [Dean et al. 1995] and RTA [Bacon and Sweeney 1996]) but they are imprecise; they abstract program executions to consider more paths than those feasible in the program. Therefore, they *introduce spurious nodes and edges, rendering large call graphs*. Relying on such algorithms for downstream analyses (*e.g.*, vulnerability detection) makes the analysis imprecise, resulting in a high amount of false positives.

A recent line of work [Santos et al. 2021, 2020], presented an approach (named Salsa) for providing support for serialization-related features. Although Salsa aids the static analyses of programs that use Java's serialization/deserialization API, it is not enough to find hidden (potentially) malicious paths in the program. Salsa relies on API modeling for abstracting the serialization/deserialization protocol which dictates callback methods control and data flow. Specifically, it relies on *downcasts* in the program to infer the callbacks invoked during **deserialization**. However, malicious objects often violate downcasts and are crafted in such way that it triggers the exploit *during* deserialization, *i.e.*, the exploit executes before the downcast is performed [Dietrich et al. 2017a].

Therefore, we introduce in this paper Seneca, a novel approach that handles the challenge of constructing call graphs for programs that use serialization features. Specifically, we are focusing on improving the call graph's *soundness* for Java programs with respect to serialization and deserialization callbacks without greatly affecting its precision. Seneca performs a ***novel taint-based call graph construction, which relies on the taint state of variables when computing possible dispatches for callback methods***.

The contributions of this work are:
— a novel taint-based call graph construction algorithm to improve *call graphs' soundness with respect to deserialization callbacks*. It is agnostic to the underlying pointer analysis method used to construct a call graph, and it is meant to complement them.
— an evaluation of the approach's soundness, precision, and scalability. Our experiments demonstrated that our approach soundly handled all the six different callbacks that can be invoked during serialization or deserialization.
— a publicly available implementation of Seneca[1]

The rest of this paper is organized as follows: Section 2 describes the serialization and deserialization mechanism and the challenges in creating a call graph that is sound with respect to this feature. Section 3 explains our approach. Subsequently, Section 4 presents the evaluation of the approach whereas Section 5 presents the results. Section 6 contextualizes our approach within the state-of-the art. Section 7 elaborates on the threats to the validity of this work. Section 8 concludes the paper and makes final considerations.

---
[1] The scripts to reproduce the paper results and Seneca's implementation are available on our GitHub repository https://github.com/s2e-lab/seneca/ and Zenodo https://zenodo.org/doi/10.5281/zenodo.10464129.

## 2 BACKGROUND

Multiple programming languages (*e.g.*, Ruby, Python, PHP, and Java) allow objects to be converted into an *abstract representation*, a process called **object serialization** (or "marshalling"). The process of reconstructing an object from its underlying abstract representation is called **object deserialization** (or "*unmarshalling*"). Serialization and deserialization of objects are widely used for inter-process communication and for improving the codes' performance by saving objects to be reused later (*e.g.*, saving machine learning models [Ten 2023]).

During object serialization/deserialization, methods from the objects' classes may be invoked. For instance, classes' constructors, getter/setter methods, or methods with specific signatures may be invoked when reconstructing the object. These are the **callback methods** of the serialization/deserialization mechanism. Each programming language has their own object serialization and deserialization protocol, abstract representation, and callback methods. The Java's default serialization and deserialization protocol is thoroughly described at their specification page [Oracle 2010]. We briefly present this mechanism in the next subsection.

### 2.1 Java Serialization API

The default Java's Serialization API converts a snapshot of an object graph into a *byte stream*. During this process, only *data* is serialized (*i.e.*, the object's fields) whereas the code associated with the object's class (*i.e.*, methods) is within the classpath of the receiver [Schneider and Muñoz 2016]. All *non-transient* and *non-static* fields are serialized by default.

The classes `ObjectInputStream` and `ObjectOutputStream` can be used for deserializing and serializing an object, respectively. They can only serialize/deserialize objects whose class implements the `java.io.Serializable` interface. If implemented by a `Serializable` class, the methods listed below can be invoked by Java ***during*** object serialization and/or deserialization:

- **`void writeObject(ObjectOutputStream)`**: it customizes the serialization of the object's state.
- **`Object writeReplace()`**: this method replaces the actual object that will be written in the stream.
- **`void readObject(ObjectInputStream)`**: it customizes the retrieval of an object's state from the stream.
- **`void readObjectNoData()`**: in the exceptional situation that a receiver has a subclass in its classpath but not its super class, this method is invoked to initialize the object's state.
- **`Object readResolve()`**: this is the inverse of `writeResolve`. It allows classes to replace a specific instance that is being read from the stream.
- **`void validateObject()`**: it validates an object after it is deserialized. For this callback to be invoked, the class has to also implement the `ObjectInputValidation` interface and register the validator by invoking the method `registerValidation` from the `ObjectInputStream` class.

Figures 1 and 2 depict the sequence of these callback method invocations. As depicted in this figure, during *serialization* of an object, the callback methods *writeReplace* and *writeObject* are invoked (if these are implemented by the class of the object being deserialized). Similarly, during object *deserializaton*, four callback methods can be invoked, namely, *readObject*, *readObjectNoData*, *readResolve*, and *validateObject*.

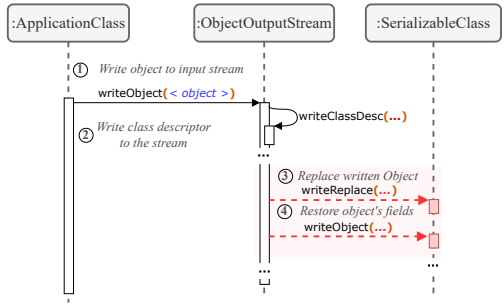

Fig. 1. Callbacks invoked during serialization

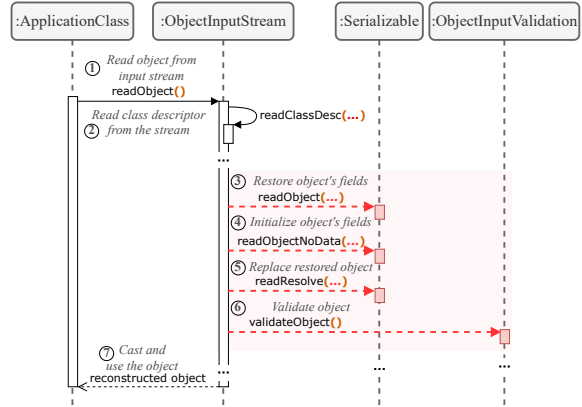

Fig. 2. Callbacks invoked during deserialization

```
1  class Pet implements Serializable {
2    protected String name;
3  }
4  class Cat extends Pet{
5    private void readObject(ObjectInputStream s){
6      /* ... */
7    }
8    private void writeObject(ObjectOutputStream s){
9      /* ... */
10   }
11 }
12 class Dog extends Pet{
13   private Object readResolve(){ /* ... */ }
14   private Object writeReplace(){ /* ... */ }
15 }
16 class Shelter implements Serializable{
17   private List<Pet> pets;
18 }
```

```
19 class SerializationExample{
20   public static void main(String[] args) throws Exception {
21     Shelter s1 = new Shelter(Arrays.asList(new Dog("Max"),
22                                            new Cat("Joy")));
23     File f = new File("pets.txt");
24     FileOutputStream fos = new FileOutputStream(f);
25     ObjectOutputStream out = new ObjectOutputStream(fos);
26     out.writeObject(s1);
27   }
28 }
29 class DeserializationExample{
30   public static void main(String[] args) throws Exception {
31     File f = new File("pets.txt");
32     FileInputStream fs = new FileInputStream(f);
33     ObjectInputStream in = new ObjectInputStream(fs);
34     Shelter s2 = (Shelter) in.readObject();
35   }
36 }
```

Listing 1. Object serialization and deserialization example

*Demonstrative Example.* Listing 1 has three serializable classes[2]: Dog, Cat and Shelter. Two of these classes have serialization callback methods (lines 5-10 and 13-14). The code at lines 21-26 serializes a Shelter object s1 into a file, whose path is provided as program arguments. The code instantiates a FileOutputStream and passes the instance to an ObjectOutputStream's constructor during its instantiation. Then, it calls writeObject(s1), which serializes s1 as a byte stream and saves it into a file. Since the object s1 has a list field (pets) that contains two objects (a Cat and a Dog instance) the callback methods of these classes invoked.

The main method at line 30 deserializes this object from the file. It creates an ObjectInputStream instance and invokes the method readObject(), which returns an object constructed from the text file. The returned object is casted to the Shelter class type. During the deserialization, the methods readObject and readResolve from the Cat and Dog classes are invoked, respectively.

*Untrusted Object Deserialization.* To illustrate how a seemingly harmless mechanism can lead to serious vulnerabilities, consider the case that the program in Listing 1 contains two more serializable classes (CacheManager and Task), as shown in Listing 2. An attacker would create a CacheManager object (cm) as shown in Figure 3. Then, the attacker serializes and encodes this malicious object (cm) into a text file and specifies it as a program argument for the main method in Listing 1. When the program reads the object from the file, it triggers the chain of method calls depicted in Figure 3. This sequence of method calls ends in an execution sink (Runtime.getRuntime.exec() on line 8 of the Task class in Listing 2).

---

[2]We only show their fields and callback methods due to space constraints.

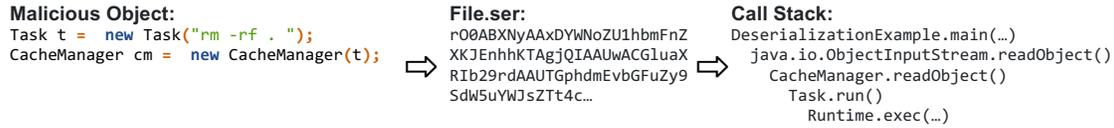

Fig. 3. Malicious serialized object used to trigger a remote code execution

```
1 public class CacheManager implements Serializable {
2   private Runnable initHook;
3   public CacheManager(Runnable initHook) {
4     this.initHook = initHook;
5   }
6   private void readObject(ObjectInputStream ois) {
7     ois.defaultReadObject();
8     initHook.run();
9   }
10 }
```

```
1 public class Task implements Runnable, Serializable {
2   private String command;
3
4   public Task(String command) {
5     this.command = command;
6   }
7   public void run() {
8     Runtime.getRuntime().exec(command);
9   }
10 }
```

Listing 2. Gadget classes that can be used to exploit an untrusted object deserialization vulnerablity

Although this request with a malicious serialized object results in a ClassCastException, the malicious command will be executed anyway, because the type cast check occurs **after** the deserialization process took place. As we can see from this example, classes can be specially combined to create a chain of method calls. These classes are called *"gadget classes"* as they are used to bootstrap a chain of method calls that will end in an execution sink.

### 2.2 Challenges for Call Graph Construction

From the examples shown in Section 2.1, we observe two major challenges that should be handled by a static analyzer in order to construct a sound call graph with respect to serialization-related features: (i) the **callback methods** that are invoked during object serialization/deserialization; and (ii) the **fields within the class can be allocated in unexpected ways**, and they dictate which callbacks are invoked at runtime. For instance, if the code snippet in Listing 1 had only the cat object in the list (line 22), then the calls to readResolve/writeReplace methods in Dog would not be made.

Existing pointer analysis algorithms leverage allocation instructions (*i.e.*, new T()) within the program to infer the possible runtime types for objects [Bastani et al. 2019; Feng et al. 2015; Heintze and Tardieu 2001; Hind 2001; Kastrinis and Smaragdakis 2013; Lhoták and Hendren 2006; Rountev et al. 2001; Smaragdakis and Kastrinis 2018]. However, as we demonstrated in the examples, the allocations of objects and their fields and invocations to callback methods are made on-the-fly by Java's serialization/deserialization mechanism. During static analysis, we can only pinpoint that there is an InputStream object that provides a stream of bytes from a source (*e.g.*, a file, socket, *etc.*) to an ObjectInputStream instance, but the contents of this stream are uncertain. Hence, the deserialized object and its state are unknown (*i.e.*, the allocations within its fields). As a result, existing static analyses fail to support serialization-related features.

## 3 SENECA: TAINT-BASED CALL GRAPH CONSTRUCTION FOR OBJECT DESERIALIZATION

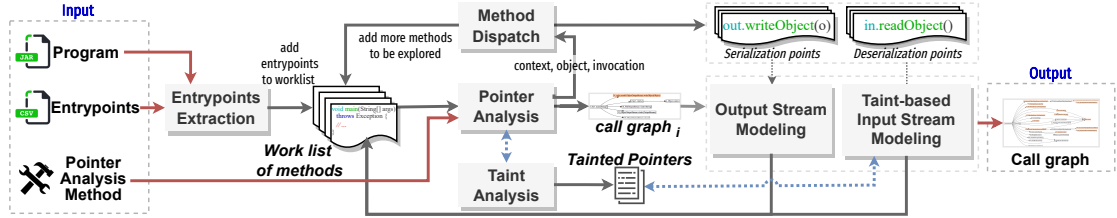

Fig. 4. Our serialization-aware approach for constructing call graphs (Seneca)

To support serialization-related features, Seneca employs an on-the-fly iterative call graph construction technique [Grove et al. 1997], as depicted in Figure 4. It involves two major phases: ① Iterating over a worklist of methods to create the *initial call graph* using an underlying pointer analysis method; ② Refinement of the initial call graph by making a set of assumptions performed iteratively until a fixpoint is reached (*i.e.*, when there are no more methods left in the worklist to be visited).

### 3.1 Phase 1: Initial Call Graph Construction

Seneca first takes as input a CSV file with method signatures for the program's **entrypoints**, which are the methods that start the program's execution (*e.g.*, main()). The result of this step is a set of entrypoint methods $m \in E$ added to our **worklist** $\mathcal{W}$. This worklist tracks the methods $m$ under a context $c$ that have to be traversed and analyzed, *i.e.*, $\langle m, c \rangle \in \mathcal{W}$, where a **context** $c$ is an abstraction of the program's state. Since the worklist $\mathcal{W}$ tracks methods within a context, the entrypoints methods added to $\mathcal{W}$ are assigned a global context, which we denote as $\emptyset$. Hence, the worklist is initialized as:

$$\mathcal{W} = \{\langle m, \emptyset \rangle \mid \forall m \in E\}$$

Starting from the entrypoint methods identified, Seneca constructs an **initial (unsound) call graph** (*i.e.*, call graph$_0$) using the underlying pointer analysis algorithm selected by the client analysis (*e.g.*, n-CFA). Each method in the worklist $\langle m, c \rangle \in \mathcal{W}$ is converted into an Intermediary Representation (IR) in Static Single Assignment form (SSA) [Cytron et al. 1991]. Each instruction in this IR is visited following the rules by the underlying pointer analysis algorithm [3]. When analyzing an instance method invocation instruction (*i.e.*, x = o.g($a_1, a_2, \ldots, a_n$)), Seneca computes the possible dispatches (call targets) for the method $g$ as follows: targets = $dispatch(pt(\langle o, c \rangle), g)$. This dispatch mechanism takes into account the current points-to set for the object $o$ at the current context $c$ as well as the declared target $g$. If the invocation instruction occurs at a **serialization** or **deserialization point**, then the *dispatch* function implemented by our approach creates a **synthetic method** to model the runtime behavior for the readObject() and writeObject() from the classes ObjectInputStream and ObjectOutputStream, respectively.

These synthetic models are initially created *without* instructions. Their instructions are constructed during the call graph refinement phase (Phase 2). It is important to highlight that the calls to synthetic methods (models) are *1-callsite-sensitive* [Sridharan et al. 2013]. We use this context-sensitiveness policy to account for the fact that one can use the same ObjectInputStream/ObjectOutputStream instance to read/write multiple objects. Thus, we want to disambiguate these paths in the call graph.

---
[3]We point the reader to the work by Sridharan *et al.* [Sridharan et al. 2013] which provides a generic formulation for multiple points-to analysis policies.

As a result of this first iteration over Phase 1, we obtain the **initial call graph** ($g_0$) and a **list of the call sites at the serialization and deserialization points**.

## 3.2 Phase 2: Call Graph Refinement

In this phase, we take as input the current call graph $g_i$ which contains as nodes actual methods in the application and synthetic methods created by our approach in the previous phase.

*3.2.1 Object Serialization Abstraction.* Algorithm 1 indicates the procedure for modeling object serialization. For each instruction at the serialization points, we obtain the points-to set for the object $o_i$ passed as the first argument to writeObject(Object). The points-to set $pt(\langle o_i, c \rangle)$ indicates the set of allocated types $t$ for $o_i$ under context $c$. Since the writeObject's argument is of type Object, we first add to $m_s$ a type cast instruction that refines the first parameter to the type $t$. In case the class type $t$ implements the writeObject(ObjectInputStream) callback, we add an invocation instruction from $m_s$ targeting this callback method.

Subsequently, we iterate over all non-static fields $f$ from the class $t$ and compute their points-to sets (see the **foreach** in line 10). If the concrete types allocated to the field contain callback methods, we add three instructions: (i) an instruction to get the instance field $f$ from the object; (ii) a downcast to the field's type; (iii) an invocation to the callback method from the field's declaring class.

It is important to highlight the edge case scenario when the type of the object being serialized is a *java.util.Collection* or a *java.util.Map*. In this case, SENECA tracks what objects were *added* to the collection in order to add invocations to their callback methods (if provided).

After adding all the needed instructions to the synthetic method $m_s$, we re-add the synthetic method to SENECA's worklist (as depicted in Figure 4).

---

**Algorithm 1:** Object serialization modeling

**Input:** Set of invocation instructions to writeObject: $I$;
Project's initial call graph: $G$;
**Output:** Set of refined synthetic models $M_s$

```
1  foreach instruction in I do
2      o_i ← argument(1, instruction)
3      c ← context(instruction)
4      m_s ← target(instruction)
5      foreach t ∈ pt(⟨o_i, c⟩) do
6          addTypeCast(m_s, t)
7          if t has a callback method then
8              addInvoke(m_s, t.callback)
9          end
10         foreach f ∈ fields(t) do
11             foreach fieldType ∈ pt(⟨o_i.f, c⟩) do
12                 if fieldType has callback then
13                     addGetField(m_s, f)
14                     addTypeCast(m_s, fieldType)
15                     addInvoke(m_s, fieldType.callback)
16                 end
17             end
18         end
19     end
20     addToWorkList(m_s, c)
21 end
```

Table 1. Taint propagation rules employed by Seneca when building call graphs.

| Instruction at method $m$ in a context $c$ | Taint propagation Rule | |
|---|---|---|
| `x = T.f` | $\tau(x) = \tau(x) \vee \tau(T.f)$ | [Load-Static] |
| `x = y.f` | $\tau(x) = \tau(x) \vee \tau(y) \vee \tau(y.f)$ | [Load-Instance] |
| `x.f = y` | $\tau(x.f) = \tau(x.f) \vee \tau(y)$ | [Store-Instance] |
| `T.f = y` | $\tau(T.f) = \tau(T.f) \vee \tau(y)$ | [Store-Static] |
| `x = o.g(a`$_1$`,`$\cdots$`,a`$_n$`)` | $\forall a_i \in A_j, \forall p_i \in P_g : \tau(p_i) = \tau(p_i) \vee \tau(a_i), \tau(g_{this}) = \tau(g_{this}) \vee \tau(o)$ | [Instance-Call-Args] |
| | $\tau(x) = \tau(x) \vee \tau(g_{ret})$ | [Instance-Call-Return] |
| | **Side Effect:** $\tau(o) = true \rightarrow pt(\langle o, c \rangle) = pt(\langle o, c \rangle) \cup targetTypes(o, c, g)$ | [Call-Side-Effect] |
| `x = T.g(a`$_1$`,`$\cdots$`,a`$_n$`)` | $\forall a_i \in A_j, \forall p_i \in P_g : \tau(p_i) = \tau(p_i) \vee \tau(a_i)$ | [Static-Call-Args] |
| | $\tau(x) = \tau(x) \vee \tau(g_{ret})$ | [Static-Call-Return] |
| `return x` | $\tau(m_{ret}) = \tau(m_{ret}) \vee \tau(x)$ | [Return] |
| | **Side Effect:** $\mathcal{W} = \mathcal{W} \cup C_m$ | [Return-Side-Effect] |
| `x = y[i]` | $\tau(x) = \tau(x) \vee \tau(y)$ | [Array-Load] |
| `x[i] = y` | $\tau(x) = \tau(x) \vee \tau(y)$ | [Array-Store] |
| `φ = v`$_1$`,v`$_2$`,`$\cdots$`,v`$_n$ | $\tau(\phi) = \tau(v_1) \vee \tau(v_2) \vee \ldots \vee \tau(v_n)$ | [Phi] |
| `x = (TypeCast) y` | $\tau(x) = \tau(x) \vee \tau(y)$ | [Checkcast] |

*3.2.2 Taint-Based Object Deserialization Abstraction.* Starting from the **deserialization points** identified, Seneca computes the call graph on-the-fly by iteratively solving constraints over the instructions. Each method in the worklist $\langle m, c \rangle \in \mathcal{W}$ is converted into an Intermediary Representation (IR) in Single Static Assignment form (SSA) [Cytron et al. 1991; Rosen et al. 1988]. Moreover, these methods have special variables to denote their return value $m_{ret}$ and the this pointer $m.this$ (for non-static methods).

For each method in the worklist $\mathcal{W}$, Seneca performs *pointer analysis* in parallel with *taint analysis* to compute the taint state of variables and points-to sets. Each instruction in the method's IR is visited following the rules by the underlying pointer analysis [Sridharan et al. 2013] and our taint analysis algorithm. Thus, each pointer in a program has an associated taint state $\tau(p)$, where $\tau(p) = true$ denotes a tainted pointer and $\tau(p) = false$ denotes an untainted (safe) pointer. Below, we provide the formulation of our taint analysis policy [Schwartz et al. 2010].

*Taint Introduction.* As described before, deserialization points are replaced by a *synthetic method*, *i.e.*, a "fake call graph node" [Sridharan et al. 2013]. It is a synthetic method created on-the-fly to model: (i) the instantiation of the class $G_c$ that contains a callback method(s) $m_c$; (ii) the invocation to the callback method(s) using the newly created object; and (iii) the instantiation of any parameters for the magic methods. It is important to highlight that in the Step (i), when instantiating the callback method's object, we invoke the class' default constructor. This is to follow the Java's deserialization process (see Section 2).

Therefore, Seneca initializes the following pointers as tainted:

- The pointer for x in the instruction `x = new `$G_c$`()`, where $G_c$ denotes a class that contains a deserialization callback method (*e.g.*, readResolve):

$$\tau(x) = \text{true}$$

- The pointers for all the fields of x:

$$\forall f_i \in fields(x) : \tau(x.f_i) = \text{true}$$

- The this pointer in the callback method $m_c$ that is invoked:

$$\tau(m_c.this) = \text{true}.$$

*Taint Propagation Rules.* As the method's instructions are parsed, we employ the rules listed in Table 1 to compute the taint states of the program's variables. As shown in Table 1, the rules for assignment instructions are as follows:

$$\text{lhs = rhs} \longrightarrow \tau(lhs) = \tau(lhs) \vee \tau(rhs)$$

That is, the pointer for the left-hand side is tainted if the pointer for the right-hand side is also tainted (or the left-hand side itself was already previously tainted). This is the case for the rules LOAD-STATIC, LOAD-INSTANCE, STORE-INSTANCE, STORE-STATIC, STATIC-CALL-RETURN, RETURN, ARRAY-LOAD, ARRAY-STORE, and CHECKCAST.

Phi functions ($\phi$) are special statements that are inserted into a method's SSA form to represent possible values for a variable depending on the control flow path taken. The taint for the pointer of phi $\tau(\phi)$ will be tainted if *any* of the possible variables' pointers are tainted.

When there is a method invocation, it can either be a static invocation or an invocation to an instance method. In both cases, each passed parameter $p_i$ is assigned to the corresponding argument $a_i$ from the invoked method. Consequently, the rules INSTANCE-CALL-ARGS, and STATIC-CALL-ARGS are propagated likewise assignment instructions. Notice, however, that for instance methods, there is a special variable $m_{this}$ denoting the "this" pointer for that method. Hence, the rule INSTANCE-CALL-ARGS propagates the taint from the caller object to the "this" pointer $\tau(g_{this})$.

It is important to highlight that **taint is never removed** from a pointer. Although this can make the underlying call graph more imprecise, our goal is to soundly reason over *all* possible runtime paths.

— Side Effects to the Pointer Analysis Engine: Method invocations and return instructions introduce *side-effects* to the static analysis engine state, labelled in Table 1 as CALL-SIDE-EFFECT and RETURN-SIDE-EFFECT, respectively.

- *Instance method invocations (CALL-SIDE-EFFECT)*: When there is an instance method invocation $o.g(...)$ and the object $o$ is tainted, then SENECA computes the possible method targets for the call $o.g(...)$ *soundly*. The dispatch is computed as described below:
  (1) it obtains the static type $t$ for $o$, i.e. $t = type(o)$;
  (2) it extracts the set of classes based on the inheritance hierarchy for $T$ (i.e., $T = cone(t)$, where $cone(t)$ returns the list of all descendants of $t$, including $t$ itself [Tip and Palsberg 2000]).
  (3) it computes the subset $C \subseteq T$ that includes only the types (classes) which provide a concrete implementation matching the signature of the invoked method $g$.
  (4) it computes the subset $A_t \subseteq C$ which includes only classes that are accessible to $t$ according to Java's visibility rules[4].
  (5) finally, the possible target methods are all the methods from the set $A_t$ in which their classes are serializable (i.e., implements the serializable interface directly or via inheritance).
  As one can notice, this dispatch is similar to the one employed by Class Hierarchy Analysis (CHA). The main difference are in steps (4) and (5), where SENECA takes into account class visibility rules as well as whether the type is serializable.
  Once the dispatch is computed (targetTypes(o,g) in CALL-SIDE-EFFECT) the points to set for $pt(\langle o, c \rangle)$ adds all the elements from $targetTypes(o, g)$.
- *Method return values (RETURN-SIDE-EFFECT)*: In a scenario where a method $m$ has a tainted return value $\tau(m_{ret}) = \text{true}$, all the callers of $m$ are re-added to the $\mathcal{W}$. Since the return is tainted, we need to back propagate this information to all the callers of $m$ to ensure that the rules INSTANCE-CALL-RETURN and STATIC-CALL-RETURN are applied correctly.

---

[4]Visibility rules are thoroughly described in the language specification https://docs.oracle.com/javase/specs/jvms/se7/html/jvms-4.html#jvms-4.1-200-E.1

— Context-sensitivity for Tainted Method Calls: Our taint-based call graph construction algorithm is ***agnostic*** to the pointer analysis policy (*e.g.*, k-l-CFA, n-CFA, *etc.*[5]). This means that a client analysis could choose to use a context insensitive analysis (*e.g.*, 0-CFA). Since tainted pointers are likely to have a large points-to set because we use a sound analysis to compute all possibilities for method dispatches when the receiver object is tainted, we should avoid merging point-to-sets of these tainted variables. Otherwise, the resulting pointer analysis would be too imprecise to be used by downstream client analyses. Therefore, we use ***1-callsite-sensitivity*** for tainted method calls (even if we use an insensitive analysis for all the other pointers).

*Demonstrative Example.* Consider the code snippet in Listing 3. The class Main has a main method that reads an object from a file, whose path is provided as a program argument. This program contains other four classes (CacheManager, TaskExecutor, CommandTask, and Config). We demonstrate Seneca's taint-based deserialization modeling strategy considering that we selected 0-1-CFA as the main pointer analysis method.

```java
class Main {
  public static void main(String[] a)
      throws Exception {
    FileInputStream f=new FileInputStream(a[0]);
    ObjectInputStream in=new ObjectInputStream(f);
    Config obj = (Config) in.readObject();
  }
}
class CommandTask
    implements Runnable, Serializable {
  private String cmd;
  private TaskExecutor taskExecutor;
  @Override
  public void run() {
    if (!cmd.isEmpty() && taskExecutor != null)
      taskExecutor.executeCmd(cmd); /* site @24 */
  }
}
class TaskExecutor implements Serializable {
  public void executeCmd(String cmd) {
    try {
      Runtime rt = Runtime.getRuntime();
      rt.exec(cmd);
    } catch (IOException e) {  }
  }
}
class Config implements Serializable {
  private String page;
  public void readObject(ObjectInputStream ois)
      throws IOException, ClassNotFoundException {
    ois.defaultReadObject();
    Runtime rt = Runtime.getRuntime();
    rt.exec("open http://localhost/" + page);
  }
}
```

```java
class CacheManager implements Serializable {
  private Runnable task;
  private Runnable[] taskArray;
  private List<Runnable> taskList;
  private Set<Runnable> taskSet;
  private Map<String, Runnable> taskMap;
  private String os;
  private long timestamp;
  public void readObject(ObjectInputStream ois)
      throws IOException, ClassNotFoundException {
    ois.defaultReadObject();
    Runnable r;
    if(os.equals("windows") && task instanceof CommandTask){
      r = getInitHook();   /* site @32 */
      r.run();
    }else {
      r = getFromArray();
      r.run();             /* site @46 */
      r = getFromList();
      r.run();             /* site @57 */
      r = getFromSet();
      r.run();             /* site @68 */
      r = getFromMap();
      r.run();             /* site @79 */
    }
  }
  Runnable getInitHook(){ return task; }
  Runnable getFromArray() { return taskArray[0]; }
  Runnable getFromList() { return taskList.get(0); }
  Runnable getFromSet() { return taskSet.iterator().next(); }
  Runnable getFromMap() { return taskMap.get("xyz"); }
}
```

Listing 3. Walk-through example to demonstrate Seneca's approach

---

[5]k-l-CFA is a family of algorithms in which k delimits the context for how many methods in the call stack the algorithm tracks and l is the context size limit which includes the object creation site and up to l-1 previously invoked methods to reach to the creation site [Grove et al. 1997; Vitek et al. 1992]. Thus, 0-1-CFA is an algorithm that distinguishes the allocations of an object based on its allocation site only (*i.e.*, new ClassA()), and ignores the call stack when the object instantiation is made. In contrast, n-CFA is an algorithm that distinguishes the allocation of an object based on its allocation site and up to n prior methods in the call stack

Seneca first extracts the program's entrypoints, provided as part of the analysis configuration. In this example, the `Main.main(String a[])` is specified as the main method. Therefore, the Seneca's worklist is initialized as: $\mathcal{W} = \{\langle Main.main(String\ a[]), \emptyset \rangle\}$. Seneca then proceeds to iteratively compute the call graph by traversing each instruction for each method in the worklist.

There are three method invocations on `Main.main()`: two invocations to the constructors (`<init>`) of `FileInputStream` and `ObjectInputStream` classes followed by a call to the `readObject()` method from the `ObjectInputStream` class. The method invocation to `ObjectInputStream.readObject()` is replaced by Seneca with a model (synthetic) method that has the same signature, but it is initialized without any instructions. At this stage, the call graph for this program after traversing the main method looks like as shown in Figure 5. All these three call graph nodes discovered after parsing `Main.main()` are added to the worklist to be processed (*i.e.*, , `FileInputStream.<init>()`, `ObjectInputStream.<init>()`, and `ObjectInputStream.readObject()`.

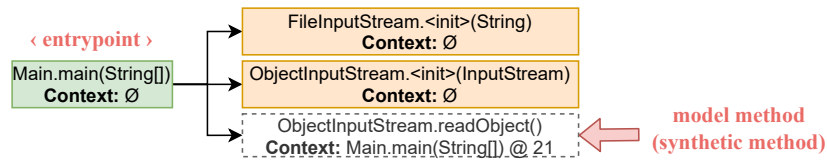

Fig. 5. Initial call graph after parsing the `Main.main()` method in Listing 3

The instructions that are added to `ObjectInputStream.readObject()` rely on taint states to infer callback methods that might be invoked during deserialization. Thus, when refining a method model, Seneca considers that *all* serializable classes in the classpath could have their callbacks invoked. By using this strategy, there are two possible callbacks that can be invoked: one from `Config` and one from `CacheManager`. Hence, all of its instance fields are marked as *tainted* per the taint introduction rules previous described (these are highlighted in red on Listing 3). Based on the taint propagation rules specified on Listing 1, variables are then marked as tainted (these variables that are tainted due to propagation are highlighted in cyan on Listing 3).

Recall that tainted invocations (*i.e.*, an instruction such as `obj.aMethod()` in which `obj` is tainted) are handled differently. Whereas the dispatch of non-tainted invocation will follow the rules from the underlying pointer analysis policy, the dispatch for tainted invocations is computed using a modified version of the CHA algorithm. Therefore, the computed call graph when using the taint-based approach looks like as Figure 6. As shown in this image[6], the model method includes the following instructions: an object instantiation for `Config` as well as `CacheManager`, their constructors' invocation, and invocations to their callback methods. Finally, the model method returns a value that can either be an instance of `Config` or `CacheManager`. Notice that the phi function ($\phi$) is added to indicate this possibility.

## 4 EVALUATION

In this section, we introduce our research questions and describe our experiment setup and design to answer those.

### 4.1 Research Questions

This paper addresses the following research questions:

**RQ1** Soundness. *Does Seneca handle object deserialization soundly?*

---

[6]For clarity, we elide the "getter" calls as well as inner calls from primordial nodes (*e.g.*, `String.isEmpty()`).

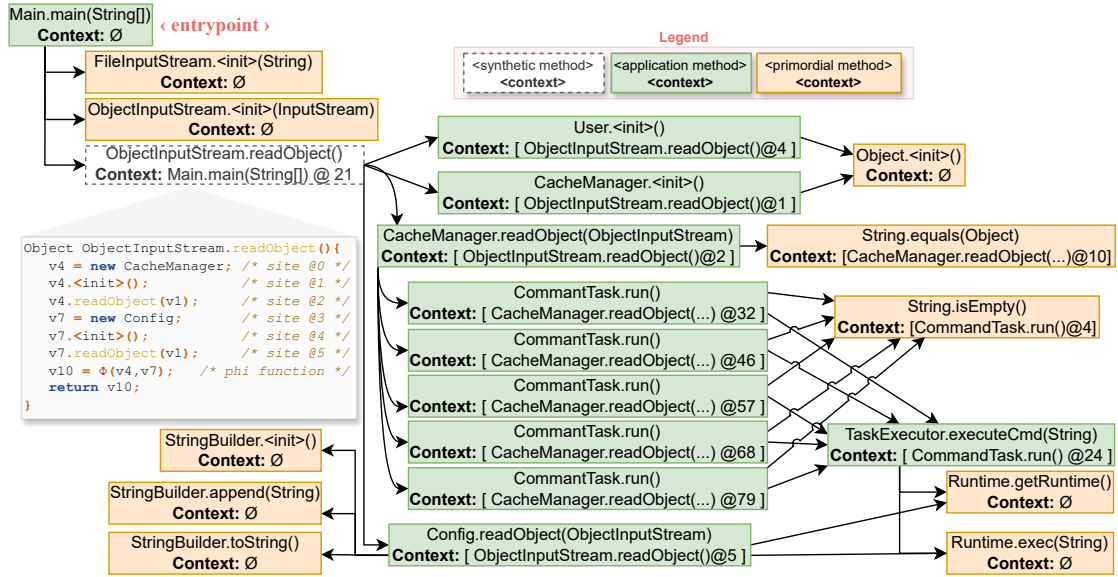

Fig. 6. Call graph for Listing 3 computed by Seneca (0-1-CFA)

**RQ2** Precision. *Does an increase in the call graph's soundness incur a significant loss in its precision?*
**RQ3** Scalability. *Does Seneca scale well for real software systems?*
**RQ4** Usefulness. *Is Seneca useful for a client analysis focused on vulnerability detection?*

To answer the aforementioned research questions, we developed a prototype for Seneca in Java using the T. J. Watson Libraries for Analysis (**Wala**) [WALA 2024]. Our prototype allows client analyses to select a pointer analysis method that can either be `0-n-CFA`, or `n-CFA`, where *n* is provided. We explain in the next subsections the methodology and datasets used to answer each RQ.

### 4.2 Answering RQ1: Soundness

We aim to verify whether Seneca improves a call graph's ***soundness*** with respect to serialization and deserialization callbacks and how it compares with existing approaches [Reif et al. 2019, 2018; Santos et al. 2021, 2020]. The soundness of a call graph construction algorithm corresponds to being able to create a call graph that incorporate ***all*** possible paths (nodes and edges) that can arise at runtime [Ali et al. 2019; Kummita et al. 2021]. In this work, we are specifically looking at improving a call graph's soundness to cover possible invocations that arise **during object serialization and deserialization**. Therefore, we use two datasets to answer this first research question.

- **Call Graph Assessment & Test Suite (CATS)** [Eichberg 2020]: This dataset was released as part of recent empirical studies [Reif et al. 2019, 2018] to investigate the soundness of the call graphs computed by existing algorithms with respect to particular programming language constructs. The CATS test suite[7] was derived by an extensive analysis of real Java projects to create test cases that are representative of common ways that projects use these language constructs (*e.g.*, lambdas, reflection, serialization, etc.). The dataset includes **9** test cases for verifying the soundness of call graphs during serialization and deserialization of objects. Each test case is a Java program with annotations that indicate the expected target for a given method call. Table 2 provides an overview of the test cases available in the CATS test suite and what aspects they aim to probe. Hence, in this first experiment, we run SENECA using two pointer analysis configurations: 0-1-CFA, and 1-CFA. Then, we compare it against SALSA (0-CFA, 1-CFA), a state-of-the-art tool, as well as the same algorithms used in the empirical study by Reif *et al.* [Reif et al. 2019], namely SOOT (CHA, RTA, VTA, and Spark), WALA (RTA, 0-CFA, 1-CFA, and 0-1-CFA), DOOP (context-insensitive), and OPAL (RTA).

---
[7]This project was formerly known as the Java Call Graph Test Suite (JCG).

Table 2. Test cases from the CATS Test Suite [Eichberg 2020] and which soundness aspect they aim to verify.

| ID | Description |
| --- | --- |
| Ser1 | The code serializes an object whose class contains a custom writeObject method. It tests whether the call graph creates a node for the writeObject(...) callback method that can be invoked by the writeObject method from the ObjectOutputStream class. |
| Ser2 | Tests whether the call graph has nodes and edges for the writeObject callback method under the scenario that the call *may* be invoked if a condition is true. |
| Ser3 | Tests whether the call graph construction algorithm considers inter-procedural flow to soundly infer that the object's writeObject(...) callback method will be invoked by the writeObject method from the ObjectOutputStream class. |
| Ser4 | The code deserializes an object (without performing a downcast) whose class contains a custom readObject method. It tests whether the call graph creates a node for the readObject(...) callback method that can be invoked by the readObject method from the ObjectInputStream class. |
| Ser5 | The code deserializes an object whose class contains a custom redObject method. It tests whether the call graph creates a node for the readObject(...) callback method that can be invoked by the readObject method from the ObjectInputStream class. Unlike Ser4, this test case has a downcast to the expected type of the read object. |
| Ser6 | Tests whether the call graph has nodes and edges for the writeReplace callback method that will be invoked during serialization. |
| Ser7 | Tests whether the call graph has nodes and edges for the readResolve callback method that will be invoked during deserialization. |
| Ser8 | Tests whether the call graph has nodes and edges for the validateObject callback method that will be invoked during deserialization. |
| Ser9 | Tests whether constructors of serializable classes are handled soundly. It checks whether the call graph models the runtime behavior, which invokes the first default constructor that is not from a serializable superclass. |

— **Metric**: Likewise to the prior empirical study by Reif *et al.* [Reif et al. 2019, 2018], we compute the number of *failed* and *passed* test cases for each approach as a way to investigate the soundness of our approach. A test case that *passes* indicates that the call graph contains the expected nodes and edges that arise during object serialization/deserialization. A *failing* test cases indicates that these nodes/edges are missing in the call graph.

• **XCorpus dataset**: Although the CATS dataset was carefully constructed to test call graph construction algorithms with respect to programming language features, the test cases are small programs (*i.e.*, with few serializable classes). There is a lack of a benchmark containing real software systems to verify how well call graph construction algorithms can handle object serialization/deserialization features Therefore, to enhance our analysis, we used programs available on the XCorpus dataset [Dietrich et al. 2017b]. We chose this dataset because it has been widely used in prior related works [Fourtounis et al. 2018; Santos et al. 2021, 2020] and it was manually curated to be representative of real Java projects.

From this dataset, we selected a total of 10 programs from the XCorpus dataset (listed in Table 3). We chose these projects because they matched the following criteria: **(i)** they perform object serialization / deserialization; **(ii)** they contain serializable classes that provide custom implementation for callback methods; hence, they would be suitable to verify whether our approach can properly compute a call graph that uncover hidden paths via callback methods.

Although the XCorpus dataset has real programs, they do not contain test cases that exercise *all* possible call paths that can go through serialization/deserialization callbacks. Thus, for each of these 10 projects, we created a set of test cases that exercised the serialization and deserialization of objects from the classes that contained custom callback methods. Each test case serializes an object into a file, and then deserializes it back from this file, as shown in Listing 4.

```java
1 public class TC<number> {
2     private static Object getObject() {
3         Object object = <initialization>
4         return object;
5     }
6     public static void main(String[] args) throws Exception {
7         Object obj = getObject();
8         FileOutputStream fOut = new FileOutputStream(args[0]);
9         ObjectOutputStream objOut = new ObjectOutputStream(fOut);
10        objOut.writeObject(obj);
11        FileInputStream fs = new FileInputStream(args[0]);
12        ObjectInputStream objIn = new ObjectInputStream(fs);
13        Object deserializedObj = objIn.readObject();
14        new File(filepath).delete();
15    }
16 }
```

Listing 4. Test Case template

The systematic process we followed to create these test cases were as follows. For each class in the XCorpus program that had a custom callback method ("gadget classes"), we created we created **5** test cases as follows:

– We created a "simple" test case. This "simple" test case returns a single instance from the class inside the method *getObject()*. That is, the object in line 3 in Listing 4 is instantiated to the type of the gadget class. We read the project's documentation to initialize the object's fields correctly and avoid exceptions thrown by the class' constructor. Moreover, we ensured that each object's reference fields are non-nulls. This will guarantee that calls to the field's class' callbacks would also show up in the runtime call graph.
– We also created "composite" test cases in which the class instance is added into a collection. These collections are an *ArrayList*, a *HashSet*, a *HashMap*, or an *array*.

By following this systematic process, we created **5** test cases (one "simple" test case, and four "composite" ones) for each class with a custom callback in an XCorpus project. We obtained a total of **210** test cases. The number of test cases per XCorpus programs is shown in Table 3.

This systematic process ensures that our test cases are sufficiently comprehensive to enable a reliable comparison of precision and soundness. With our test cases, *all* possible de/serialization-related callback methods in the program's classpath are executed at least once. Moreover, this also ensures that we test the computed call graphs with respect to handling serializable *collections* containing other serializable objects.

After creating these test cases, we execute them to extract their **dynamic call graph** (runtime call graph). We implemented a JVMTI (Java Virtual Machine Tool Interface) agent in C to compute these runtime call graphs. This implementation has an instrumentation agent that is attached to the program's execution. It captures every method that is invoked in the program and its caller method.

Since we aim to investigate whether our taint-based call graph algorithm handle object deserialization soundly or would unsound assumptions be able to find vulnerabilities, we compare Seneca against Salsa [Santos et al. 2021, 2020], a state-of-the-art approach that computes call graphs for object deserialization based on downcasts within the program, which yields to less sound call graphs.

**Metric**: Similar to prior works [Ali et al. 2019; Ali and Lhoták 2012; Kummita et al. 2021; Li et al. 2014; Smaragdakis et al. 2015], we verify our approach's soundness based on the number of edges in the runtime call graph that are *missing* in the static call graph. In our comparison, we differentiate *application-to-application* edges, *application-to-library*, and *library-to-library* edges. That means that we disregard missing edges due to: (a) *class initializers* (because `<clinit>` methods are modeled by Wala using a synthetic method that invokes all class initializers at once), (b) *native code* (because it cannot be statically analyzed), (c) *explicitly excluded classes* (*i.e.*, classes inside our list of exclusions file that are removed from the static call graph), and (d) *library-to-library* edges (*i.e.*, edges from a built-in Java class to another built-in language class).

Table 3. XCorpus programs [Dietrich et al. 2017b] used in our experiments and the number of test cases created for each of them.

| Project | batik (1.7) | castor (1.3.1) | james (2.2.0) | jgraph (5.13.0.0) | jpf (1.5.1) | log4j (1.2.16) | openjms (0.7.7-beta-1) | pooka (3.0-080505) | xalan (2.7.1) | xerces (2.10.0) |
|---|---|---|---|---|---|---|---|---|---|---|
| # Classes | 2,560 | 1,639 | 340 | 187 | 152 | 308 | 808 | 1,617 | 1,621 | 1,034 |
| # Classes in Dependencies | 1,209 | 947 | 274 | 0 | 1 | 0 | 28 | 0 | 0 | 0 |
| # Test Cases | 25 | 65 | 5 | 30 | 5 | 15 | 5 | 30 | 25 | 5 |

### 4.3 Answering RQ2: Precision

Although soundness is a desirable property for static analysis, in practice, however, creating a sound analysis also implies a loss of precision. Due to the undecidability of program verification, it is impossible to create an analysis that is both *sound* and *precise* [Rice 1953]. Therefore, a sound analysis is an over-approximation that may include spurious results (*e.g.*, unrealistic paths).

While our approach aims to enhance an existing call graph construction algorithm to handle serialization-related callbacks soundly, we need to verify whether our approach introduces imprecision and to what extent. Imprecision in this work refers to adding nodes and edges that will not arise at the program's runtime during object serialization and deserialization [Ali et al. 2019].

To answer this question, we use our JVMTI agent to compute the runtime call graph for each program in the CATS test suite [Eichberg 2020] and our manually constructed test cases derived from the XCorpus dataset [Dietrich et al. 2017b]. Subsequently, we compute the number of edges in the *static* call graph that did not exist in the *runtime* call graph.

— **Metric**: We calculate the number of nodes and edges that appeared in Seneca's call graph but did not appear on the dynamic call graph. Similar to prior works [Smaragdakis et al. 2015; Smaragdakis and Kastrinis 2018], when performing this calculation, we only consider *application-to-application* edges and *application-to-library* edges as long as these edges do not include nodes that are a *class initializer*, a *native code method*, or a method from an *explicitly excluded class*.

### 4.4 Answering RQ3: Performance

Our serialization-aware call graph construction approach introduces *extra iterations* on the underlying pointer analysis methods. As such, we investigate whether these extra iterations introduce significant overhead that renders the analysis impractical for real large-scale programs.

To verify the overhead of incurred by our approach, we first use Seneca to build the call graphs for the test cases created for the 10 programs extracted from the XCorpus dataset [Dietrich et al. 2017b]. Subsequently, we run the `0-1-CFA` and `1-CFA` call graph construction algorithms available in Wala **with** and **without** our serialization-aware approach enabled. For comparison, we also ran Salsa configured with `0-1-CFA` and `1-CFA` to build call graphs. For all of these approaches, we used a standard list of class exclusions[8]; these classes are ignored during call graph construction by Wala, Salsa, and Seneca.
— *Metric*: We measure **(i)** the *running time* to compute the call graphs when using our approach, and **(ii)** the extra *added number of iterations* over the worklist of the call graph construction algorithm. We run these analyses on a machine with a 2.9 GHz Intel Core i7 processor and 32 Gb of RAM memory.

### 4.5 Answering RQ4: Efficiency

One of the premises of this work is that a taint-based call graph construction enables the computation of sound call graphs with respect to (de)serialization, which can be useful for client analyses, such as vulnerability detection. In this question, we aim to verify whether Seneca can help a static analysis technique in finding potential vulnerable paths in the program.

To answer this question, we obtained 3 open-source projects with known disclosed deserialization vulnerabilities. We selected these projects because their exploits have been widely discussed by practitioners and are available on the YSoSerial GitHub repository [Frohoff 2018]. That is, these projects have well-known "gadget chains" which were previously disclosed in vulnerability reports.

To answer this RQ, we used Seneca and Salsa to compute the call graphs of these projects. Each technique was configured to use 0-1-CFA and 1-CFA. Subsequently, we use these call graphs to extract vulnerable paths which are paths from ObjectInputStream.readObject() to *sinks*, *i.e.*, method invocations to security-sensitive operations.

To identify sinks, we manually curated a list of security-sensitive method signatures. To do so, we extracted the list of sink methods from a prior published work [Thomé et al. 2017]. Moreover, we parsed the manifest file from the Juliet Test Suite [NSA Center for Assured Software 2017]. This test suite is a dataset from NIST (National Institute of Standards and Technology) which has a collection of synthetic C/C++ and Java code samples with different software weaknesses (CWEs). Their manifest file indicates all the files for a test case, the kind of weakness it contains, and its location in the code. Thus, we parsed the manifest to extract the lines that are flagged as vulnerable, filtered out the lines that are not method invocations, grouped them by signature, and manually identified the ones that are sinks. After performing these two complementary curation steps, we obtained a total of **101** methods signatures for sinks.
— *Metric:* We measured how many *vulnerable paths* each approach was able to identify.

## 5 RESULTS

### 5.1 RQ1: Call Graph Soundness

This section describes the results of the experiments for measuring the soundness of the call graphs computed by Seneca.

*5.1.1 Dataset #1: CATS.* Table 4 reports the programs in which each approach soundly inferred the call graph (✓) and the ones it failed to do so (✗). As shown in this table, we built call graphs using two different pointer analysis policies: 0-1-CFA, and 1-CFA. For the sake of comparison, this table also includes the same algorithms and results presented by Reif *et al.* [Reif et al. 2019] and that we were able to reproduce using the Docker image [Reif 2023] provided by their work. The released artifacts of Reif *et al.* study [Reif et al. 2019] includes adapters for constructing call graphs using Soot (CHA, RTA, VTA, and Spark), Wala (RTA, 0-CFA, 1-CFA, and 0-1-CFA), and Opal (RTA). We also included a comparison with a recent published work, Salsa [Santos et al. 2021, 2020], configured with the same pointer analysis policies as ours, *i.e.*, 0-1-CFA, and 1-CFA.

---

[8]https://github.com/wala/WALA/blob/master/com.ibm.wala.core/src/main/resources/Java60RegressionExclusions.txt

Table 4. Passed/failed test cases from CATS

| | Salsa (0-1-CFA) | Salsa (1-CFA) | Seneca (0-1-CFA) | Seneca (1-CFA) | Opal (RTA) | Soot (CHA) | Soot (RTA) | Soot (Spark) | Soot (VTA) | Wala (0-1-CFA) | Wala (0-CFA) | Wala (1-CFA) | Wala (RTA) |
|---|---|---|---|---|---|---|---|---|---|---|---|---|---|
| Ser1 | ✓ | ✓ | ✓ | ✓ | ✓ | ✗ | ✗ | ✗ | ✗ | ✗ | ✗ | ✗ | ✗ |
| Ser2 | ✓ | ✓ | ✓ | ✓ | ✓ | ✗ | ✗ | ✗ | ✗ | ✗ | ✗ | ✗ | ✗ |
| Ser3 | ✓ | ✓ | ✓ | ✓ | ✗ | ✗ | ✗ | ✗ | ✗ | ✗ | ✗ | ✗ | ✗ |
| Ser4 | ✓ | ✓ | ✓ | ✓ | ✗ | ✗ | ✗ | ✗ | ✗ | ✗ | ✗ | ✗ | ✗ |
| Ser5 | ✓ | ✓ | ✓ | ✓ | ✓ | ✗ | ✗ | ✗ | ✗ | ✗ | ✗ | ✗ | ✗ |
| Ser6 | ✓ | ✓ | ✓ | ✓ | ✗ | ✗ | ✗ | ✗ | ✗ | ✗ | ✗ | ✗ | ✗ |
| Ser7 | ✓ | ✓ | ✓ | ✓ | ✓ | ✗ | ✗ | ✗ | ✗ | ✗ | ✗ | ✗ | ✗ |
| Ser8 | ✓ | ✓ | ✓ | ✓ | ✗ | ✓ | ✓ | ✗ | ✗ | ✗ | ✗ | ✗ | ✗ |
| Ser9 | ✓ | ✓ | ✓ | ✓ | ✓ | ✓ | ✓ | ✗ | ✗ | ✗ | ✗ | ✗ | ✗ |

As shown in Table 4, only our serialization-aware call graph construction (Seneca) and Salsa passed **all** of the nine test cases. Only three other algorithms partially provided support for callback methods, namely Soot$_{RTA}$ and Soot$_{CHA}$ (2 out of 9) and Opal$_{RTA}$ (5 out of 9) [Reif et al. 2019]. The remaining algorithms, *i.e.*, Soot (VTA, and Spark), Wala (RTA, 0-CFA, 1-CFA, 0-1-CFA), did not provide support at all for serialization-related callback methods.

It is also important to highlight that the frameworks that provided partial support for serialization-related features (Soot$_{RTA}$, Soot$_{CHA}$, and Opal$_{RTA}$) use ***imprecise*** call graph construction algorithms (CHA [Dean et al. 1995] or RTA [Bacon and Sweeney 1996]). Table 5 shows a comparison of call graphs' sizes in terms of nodes and edges. As we can infer from these charts, the only call graph construction algorithms used by Soot, and Opal that provided **partial** support for serialization create much larger call graphs (in terms of the number of nodes and edges). Since these algorithms only rely on static types when computing the possible targets of a method invocation, they introduce spurious nodes and edges, thereby increasing the call graph's size.

Table 5. Call graph sizes for each approach and test case (TC) from the CATS benchmark.

| TC | Approach | # Nodes | # Edges | TC | Approach | # Nodes | # Edges | TC | Approach | # Nodes | # Edges |
|---|---|---|---|---|---|---|---|---|---|---|---|
| Ser1 | Opal$_{RTA}$ | 5,983 | 39,580 | Ser4 | Salsa$_{1-CFA}$ | 1,590 | 2,841 | Ser7 | Seneca$_{0-1-CFA}$ | 722 | 1,323 |
| | Salsa$_{0-1-CFA}$ | 771 | 1,527 | | Seneca$_{0-1-CFA}$ | 722 | 1,323 | | Seneca$_{1-CFA}$ | 1,590 | 2,841 |
| | Salsa$_{1-CFA}$ | 1,876 | 3,538 | | Seneca$_{1-CFA}$ | 1,590 | 2,841 | | Salsa$_{0-1-CFA}$ | 729 | 1,333 |
| | Seneca$_{0-1-CFA}$ | 771 | 1,527 | Ser5 | Opal$_{RTA}$ | 6,461 | 44,773 | | Salsa$_{1-CFA}$ | 1,601 | 2,855 |
| | Seneca$_{1-CFA}$ | 1,876 | 3,538 | | Salsa$_{0-1-CFA}$ | 722 | 1,323 | Ser8 | Seneca$_{0-1-CFA}$ | 729 | 1,333 |
| Ser2 | Opal$_{RTA}$ | 5,985 | 39,583 | | Salsa$_{1-CFA}$ | 1,590 | 2,841 | | Seneca$_{1-CFA}$ | 1,601 | 2,855 |
| | Salsa$_{0-1-CFA}$ | 772 | 1,529 | | Seneca$_{0-1-CFA}$ | 722 | 1,323 | | Soot$_{CHA}$ | 17,570 | 261,274 |
| | Salsa$_{1-CFA}$ | 1,878 | 3,540 | | Seneca$_{1-CFA}$ | 1,590 | 2,841 | | Soot$_{RTA}$ | 17,449 | 259,257 |
| | Seneca$_{0-1-CFA}$ | 772 | 1,529 | Ser6 | Salsa$_{0-1-CFA}$ | 546 | 940 | | Opal$_{RTA}$ | 6,463 | 44,775 |
| | Seneca$_{1-CFA}$ | 1,878 | 3,540 | | Salsa$_{1-CFA}$ | 1,068 | 1,718 | | Salsa$_{0-1-CFA}$ | 724 | 1,325 |
| Ser3 | Salsa$_{0-1-CFA}$ | 772 | 1,528 | | Seneca$_{0-1-CFA}$ | 546 | 940 | Ser9 | Salsa$_{1-CFA}$ | 1,592 | 2,843 |
| | Salsa$_{1-CFA}$ | 1,877 | 3,539 | | Seneca$_{1-CFA}$ | 1,068 | 1,718 | | Seneca$_{0-1-CFA}$ | 724 | 1,325 |
| | Seneca$_{0-1-CFA}$ | 772 | 1,528 | Ser7 | Opal$_{RTA}$ | 6,458 | 44,763 | | Seneca$_{1-CFA}$ | 1,592 | 2,843 |
| | Seneca$_{1-CFA}$ | 1,877 | 3,539 | | Salsa$_{0-1-CFA}$ | 722 | 1,323 | | Soot$_{CHA}$ | 17,570 | 261,302 |
| Ser4 | Salsa$_{0-1-CFA}$ | 722 | 1,323 | | Salsa$_{1-CFA}$ | 1,590 | 2,841 | | Soot$_{RTA}$ | 17,449 | 259,286 |

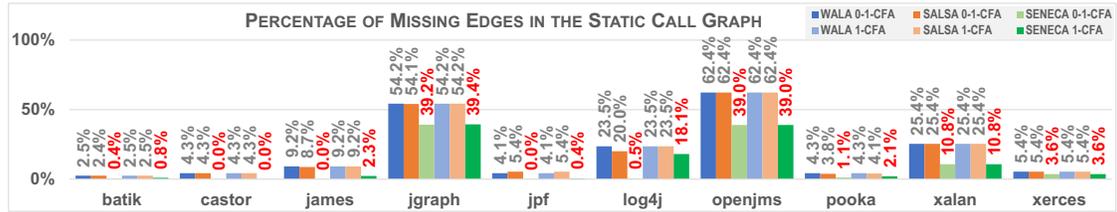

Fig. 7. Percentage of missing edges in the static call graphs computed by Wala, Salsa, and Seneca for the projects in the XCorpus dataset

Our approach enhances the underlying pointer analysis policy in order to strike a balance between improving soundness while not greatly affecting the call graph's precision by adding spurious nodes and edges. A more recent work, Salsa, also produced call graphs with reasonable sizes similar to ours. However, this similar performance is caused by the fact that the test cases in the CATS dataset are rather simple; they are up to two classes that exercise one custom call back method at a time. As we will discuss in the next subsection, Salsa's ability to create sound call graphs is greatly diminished when building the call graph for real software projects.

*5.1.2 Dataset #2: XCorpus Dataset.* Figure 7 depicts the percentage of edges in the runtime call graph of the projects, that are *missing* on the static call graph computed by each approach. From this chart, we notice that Seneca **outperformed** Wala and Salsa. Our approach has ***less*** missing edges compared to other the approaches, *i.e.*, it is able to soundly infer hidden paths through serialization callbacks.

For the *castor* project, Seneca ***did not miss any runtime edge***. In contrast, Wala and Salsa (0-1-CFA and 1-CFA) missed 4.3% of the runtime edges. Seneca$_{0-1-CFA}$ also did not miss any runtime edges for two other projects (*james*, and *jpf*), whereas Wala$_{0-1-CFA}$ and Salsa$_{0-1-CFA}$ missed 8.7% and 5.4% of edges, respectively. The biggest improvements in comparison to other approaches were observed for the test cases created for the *jgraph*, *openjms*, *log4j*, and *xalan* projects. The percentage difference between Seneca and Wala as well as Salsa ranged from 5% to 23.4%.

When inspecting the edges that Seneca missed, we observed that these edges were *unrelated* to serialization callbacks. That is, these were edges to which the underlying pointer analysis algorithm cannot soundly infer the points-to sets of variables. For example, we observed edges that were missed because instructions were using reflection to invoke methods. These were constructs that the underlying 0-1-CFA and 1-CFA pointer analysis provided by Wala (our baseline framework) could not correctly infer the dispatch.

One of the reasons as to why Salsa performed similar to Seneca with the CATS test suite but performed poorly on the XCorpus dataset has to do with its inability to compute potential method dispatches from classes in the classpath. As described in their work [Santos et al. 2021, 2020], the approach relies on downcasts of objects to infer what are the object(s) being deserialized. When downcasts are unavailable, the approach relies on a simple approach of computing all possible dispatches, but limited to classes on the application scope. Our approach, on the other hand, follows Java's serialization specification and includes *all* classes in the classpath, irrespective of its scope (*i.e.*, extension, primordial, or application scope).

**Summary of Findings for RQ1**

– Our experiments showed that our approach **improved** a call graphs' soundness with respect to serialization-related features. It added nodes and edges in the call graph that could arise at runtime during serialization and deserialization of objects.
– Our approach **passed all test cases**, whereas other approaches, namely Soot$_{RTA}$, Soot$_{RTA}$ passed only 2, and OPAL$_{RTA}$ passed 5.
– The only call graph construction algorithms used by Soot, and Opal that provided **partial** support for serialization used algorithms that only rely on the method's signatures for dispatch (*i.e.*, CHA and RTA). Hence, they created much larger call graphs because they introduced spurious nodes and edges.
– Although Salsa, a recently published work, also passed all the test cases in the CATS test suite, it failed to soundly infer the callbacks in real applications from the XCorpus dataset.

## 5.2 RQ2: Precision

This section describes the evaluation results of the precision of the call graphs computed by Seneca.

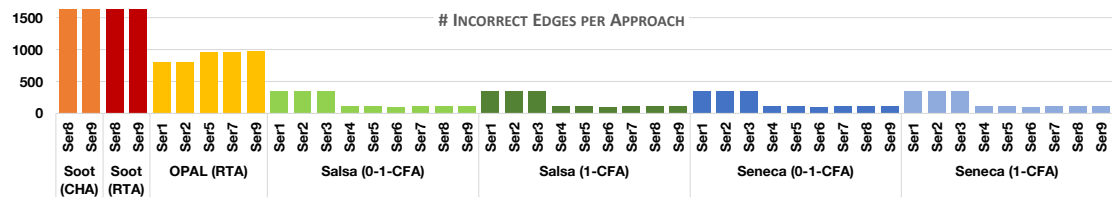

Fig. 8. Precision results for the test cases from the CATS test suite.

*5.2.1 Dataset #1: CATS.* Figure 8 depicts the number of edges in the *static* call graph that were not found in the *runtime* call graph for the test cases in the CATS test suite [Reif et al. 2019]. As shown in this chart, Seneca was able to provide full support for serialization callbacks (passing all test cases, see Table 4) while maintaining reasonably sized call graphs. Soot and OPAL derived call graphs that were far more *imprecise*. While Opal and Soot had over 800 imprecise edges (false positives - FP), Seneca had between 95 and 343 incorrect edges. Therefore, Soot and Opal had, on average, 8.8× times and 4.8× times incorrect edges than Seneca, respectively.

This comparison also shows that Salsa's performance was similar to Seneca. As explained in the previous section, however, this similar performance is caused by the fact that the programs in the CATS test suite are *small*; they do not include scenarios where Salsa's unsound assumptions fall short.

*5.2.2 Dataset #2: XCorpus Dataset.* Figure 9 plots the percentage of edges that are in the runtime call graph, but that are not in the static call graph of each approach. As observed on this chart, unsurprisingly, increasing the soundness of the call graph also increased the number of imprecise edges (*i.e.*, edges that did not arise at runtime). The increase of missed edges is comparable to the one by Salsa.

When we inspected the imprecise edges, we noticed that those were related to serialization nodes, *i.e.*, cases in which our call graph included *all* possible objects that can be serialized. Indeed, as our test cases serialized only *one* object at a time, all these edges are deemed as incorrect. However, as the Java API allows the deserialization of arbitrary types (*i.e.*, any serializable type available on the class path), the edges in Seneca could arise at runtime if an object being read uses one of the other serializable classes (other than the one from the test case).

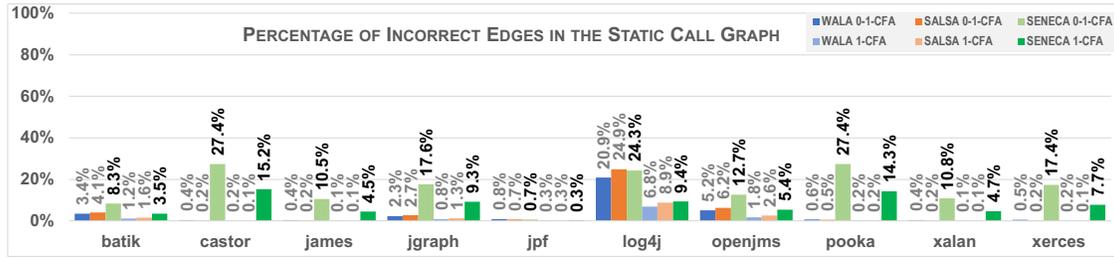

Fig. 9. Percentage of incorrect edges (*i.e.*, edges in the runtime CG not in the static CG) for each approach

### Summary of Findings for RQ2
– For the CATS dataset, Soot and Opal computed call graphs that were far more imprecise than Seneca, an average of 8.8× and 4.8× more incorrect edges, respectively.
– While Salsa and Seneca had a similar amount of incorrect edges for the CATS benchmark, Salsa produced call graphs with more imprecise edges than Seneca for the test cases created for the projects in the XCorpus dataset.

## 5.3 RQ3: Performance

We measured the running time observed when computing the call graphs using Wala, Salsa, and Seneca, configured with 0-1-CFA and 1-CFA pointer analysis policies. The results for these experiments are shown in Figure 10. As we would expect, Seneca takes longer to compute call graphs as it has to process nodes and edges related serialization.

The observed differences, however, do not hinder the overall scalability of the approach. The approach still finishes within seconds of execution. Moreover, when further inspecting the worklist of our algorithm, we noticed that Seneca incurs between 3–6 extra iterations over Wala's worklist. These extra iterations along with the taint analysis are the root cause for the extra running time needed for Seneca to finish.

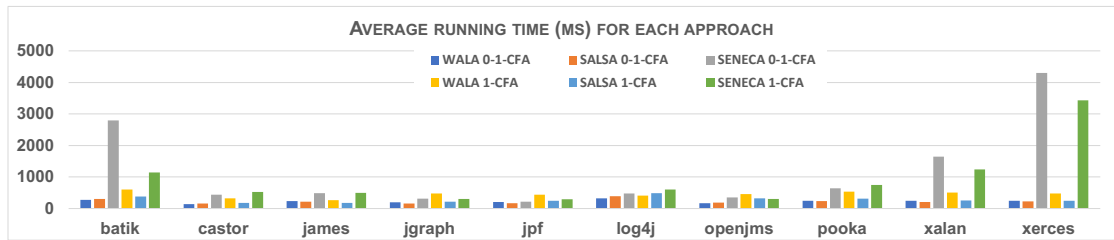

Fig. 10. The total running time (milliseconds) that it took each approach to compute a call graph.

### Summary of Findings for RQ3
– Seneca's performance, when evaluated against an established benchmark, has been found to not induce great overhead on the underlying call graph construction approach. This makes Seneca a viable option for developers and researchers in need of sound call graph for analyzing programs that heavily use serialization constructs.

## 5.4 RQ4: Usefulness for Vulnerability Detection

We have implemented a client analyses that attempts to find vulnerable paths caused by untrusted object deserialization in a program. We then verified how well this client analysis could detect vulnerable paths by comparing its performance using the call graph computed by SALSA and the one generated by SENECA. The results for this experiment are shown in Table 6.

Table 6. Number vulnerable paths found by a client analyses that used SALSA's and SENECA's call graphs

|  | FileUpload | | | | Vaadin | | | | Wicket | | | |
|---|---|---|---|---|---|---|---|---|---|---|---|---|
|  | SALSA | | SENECA | | SALSA | | SENECA | | SALSA | | SENECA | |
|  | 0-1-CFA | 1-CFA | 0-1-CFA | 1-CFA | 0-1-CFA | 1-CFA | 0-1-CFA | 1-CFA | 0-1-CFA | 1-CFA | 0-1-CFA | 1-CFA |
| # Vuln. Paths | 0 | 0 | 14 | 12 | 0 | 0 | 4 | 0 | 0 | 0 | 20 | 34 |

As shown in this table, SALSA's call graphs were not suitable for performing vulnerability detection. The key issue lies on the unsoundness of SALSA. This approach relies on type casts (downcasts) to infer what object is being deserialized from a stream. However, as explained in Section 2.1, untrusted object deserialization vulnerabilities are caused by the ability of an attacker to craft arbitrary objects using *any* serializable class available in the classpath. Thus, even if the program performs a downcast over the serialized object, the exploit would have been executed anyway, as the vulnerability arises *during* deserialization and not after it.

Unlike SALSA, our approach was able to find vulnerable paths within our allocated time budget (of 15 minutes and up to 15 call graph nodes in a path). The identified paths included the vulnerable paths from previously disclosed gadget chains, documented on the YSoSerial repository of deserialization exploits [Frohoff 2018].

> **Summary of Findings for RQ4**
> – We showed the benefits of a sound call graph with respect to deserialization by implement a client static analysis that detect vulnerable paths caused by untrusted object deserialization. Our results showed that while SENECA is able to find previously disclosed vulnerable paths, an existing approach (SALSA) falls short in generating call graphs that can infer these hidden vulnerable paths.
> – The experiments highlight the importance of building call graphs that are sound with respect to deserialization features and demonstrate that SENECA can be suitable for downstream analyses that require the handling of serialization constructs in a sound fashion.

## 6 RELATED WORK

This section discusses relevant works related to object deserialization and call graph construction.

### 6.1 Call graph Construction & Taming Challenging Programming Features

Call graphs are a core data structure for multiple analyses. Thus, previous works focused on devising algorithms for their construction. Among these works, we have CHA [Dean et al. 1995] and RTA [Bacon and Sweeney 1996], which are two well-known algorithms that over-approximates possible call paths by relying on methods' signatures. Since these algorithms are overly conservative, multiple works discussed frameworks to make them more precise [Grove and Chambers 2001; Grove et al. 1997; Tip and Palsberg 2000]. Moreover, previous research also focused on creating application-only call graphs, that disregard unnecessary library classes, while keeping on the graph the nodes and edges that are important for the underlying analysis [Ali and Lhoták 2012]. In this paper, we focused on solving the challenge of computing call graphs that are sound concerning object serialization and deserialization.

Previous research on static analysis also explored the challenges involving supporting reflection features [Bodden et al. 2011; Li et al. 2014, 2019; Smaragdakis et al. 2015], dynamic proxies [Fourtounis et al. 2018], enterprise frameworks [Antoniadis et al. 2020] and RMI-based programs [Sharp and Rountev 2006]. These approaches involve making assumptions when performing the analysis, to create analyses that are not overly imprecise. Unlike these prior works, however, we focused on object deserialization that has its own unique challenges, as described in Section 2.2.

### 6.2 Empirical Studies on Call Graphs

Multiple characteristics of call graphs (*e.g.*, precision, soundness, performance, and recall) have been widely studied in the past [Ali et al. 2019; Kummita et al. 2021; Murphy et al. 1998; Sui et al. 2020]. Murphy *et al.* [Murphy et al. 1998] studied multiple call graph construction approaches for C programs, finding discrepancies among the generated call graphs across different approaches. Sui *et al.* [Sui et al. 2018] focused on the support for dynamic language features, aiming to create a benchmark for dynamic features for Java.

There is a line of research that explored call graph's soundness of Java (or JVM-like) programs [Ali et al. 2019; Reif et al. 2019, 2018]. In particular, recent empirical studies [Reif et al. 2019, 2018] show that although serialization-related features are widely used, they are not well-supported in existing approaches. Thus, we built an approach to enhance existing points-to analysis to support the construction of sound call graphs with respect to serialization-related features.

### 6.3 Pointer Analysis

Many works explored the problem of performing pointer analysis of programs [Bastani et al. 2019; Feng et al. 2015; Heintze and Tardieu 2001; Hind 2001; Kastrinis and Smaragdakis 2013; Lhoták and Hendren 2006; Rountev et al. 2001; Smaragdakis and Kastrinis 2018]. These approaches focus on computing over- or under-approximations to improve one or more aspects of the analysis, such as its soundness, precision, performance, and scalability. Existing pointer analysis approaches make the sets finite such that the problem can be algorithmically solvable. In this paper, however, we focus on aiding points to analysis to soundly handle serialization-related features in a program, which are currently not well-supported because it relies on reflection [Reif et al. 2018].

### 6.4 Detecting Untrusted Object Deserialization

More recently there were approaches published that aimed at detecting untrusted object deserialization for PHP [Koutroumpouchos et al. 2019; Shahriar and Haddad 2016] and .NET [Shcherbakov and Balliu 2021]. Shcherbakov and Balliu [Shcherbakov and Balliu 2021] described an approach to semi-automatically detect and exploit object injection vulnerabilities .NET applications. It relies on existing publicly available gadgets to perform the detection and exploitation. Koutroumpouchos *et al.* described ObjectMap [Koutroumpouchos et al. 2019] which is tool that performs black-box analysis of Web applications to pinpoint potential insecure deserialization vulnerabilities. It works by inserting payloads into the parameters of HTTP GET/POST requests and then monitoring the target web application for errors to infer whether the application is vulnerable or not.

Recent works [Cao et al. 2023; Haken 2018; Rasheed and Dietrich 2020] focused on deserialization vulnerabilities in Java programs. Rasheed and Dietrich [Rasheed and Dietrich 2020] described a hybrid approach that first performs a static analysis of a Java program to find potential call chains that can lead to sinks, where reflective method calls are made. It then uses the results of the static analysis to perform fuzzing in order to generate malicious objects.

Unlike these prior works, we aimed to create an approach that can create sound call graphs with respect to serialization-related features. Our call graph is intended to be used by downstream client analyses, including, but not limited to, vulnerability detection.

### 6.5 Empirical Studies on Untrusted Object Deserialization

In the past few years, we observed a spike of vulnerabilities associated with deserialization of objects [Cifuentes et al. 2015]. Thus, existing works also studied vulnerabilities rooted at untrusted deserialization vulnerabilities [Dietrich et al. 2017a; Peles and Hay 2015; Sayar et al. 2023]. Peles *et al.* [Peles and Hay 2015] conducted an empirical investigation of deserialization of pointers that lead to vulnerabilities in Android applications and SDKs. Dietrich *et al.* [Dietrich et al. 2017a] demonstrated how seemingly innocuous objects trigger vulnerabilities when deserialized, leading to denial of service attacks. In this paper, we describe an approach that could help client analyses focused on detecting instances of untrusted object deserialization. Sayar *et al.* [Sayar et al. 2023] investigated deserialization vulnerabilities in Java applications, showing that a significant proportion of libraries contain unpatched exploitable code fragments (gadgets) as well as many of the studied vulnerabilities were improperly fixed or only mitigated through workarounds.

While these previous works highlighted the critical nature of untrusted object deserialization, our paper aims to create call graphs that are sound with respect to serialization/deserialization features in Java programs. Our approach can be used as a building block for downstream static analyzers to detect untrusted object deserialization in Java codebases.

### 7 THREATS TO VALIDITY

The main construct validity threat [Runeson and Höst 2009] to this work relates to how we measure *soundness* and *precision* of the constructed call graphs. To compute these metrics, we followed similar methodology employed by several prior works [Ali et al. 2019; Ali and Lhoták 2012; Kummita et al. 2021; Li et al. 2014; Smaragdakis et al. 2015]. Specifically, we measure soundness and precision by comparing *static* call graphs to *runtime* call graphs.

One of the key challenges when it comes to computing soundness and precision is that it is difficult to create a program's *ground truth*, that is, the call graph that represents *all* possible runtime behaviors at runtime. To mitigate this threat and ensure that our runtime call graphs are a suitable approximation of the ground truth, we created **210** test cases in which we had real programs with several objects being serialized/deserialized, such that we cover *all* possible serialization callbacks available in the program's classpath (as described in Section 4). Each test case is a program that serializes/deserializes one object (this object can be an individual serializable object, or a serializable collection containing another serializable object). This way, the computed runtime call graphs can enable a reliable comparison of soundness and precision of for the call graphs under evaluation.

Another threat to the validity of this work is that in RQ1 (Section 5.1, Table 5) we compare the call graphs' sizes as means to approximate how precise they are (*i.e.*, smaller call graphs are likely more precise than bigger ones). While it is possible that the smaller call graphs are missing required edges, in this particular experiment, we used a manually curated benchmark (CATS) which include *small* programs that contain up to two classes each and are enriched with annotations that indicate the expected target for a given method call. As such, since we verified whether each call graph contained the expected serialization-related callback nodes and edges and the programs are expected to be small, a bigger call graph that is missing required edges is, consequently, imprecise. In fact, while Opal and Soot did not include all the expected nodes/edges (see Table 4), they had much larger call graphs than Salsa and Seneca (Table 5).

An *external validity threat* [Runeson and Höst 2009] to this work concerns the fact that we evaluated Seneca's performance on programs that used serialization features. That is, we have not measured the performance impact on programs that do not use (or rarely use) serialization features. However, it is important to highlight that Seneca does not have an additional overhead in programs that do not use serialization. Seneca *only* incurs extra iterations on the underlying call graph construction algorithm upon encountering a serialization/deserialization point (as described in Section 3). Thus, when a program does not perform serialization/deserialization or use these sparingly, the overhead is similar as to WALA; its baseline technique.

## 8 CONCLUSION

We presented an approach to support the static analysis of serialization-related features in Java programs. It works under the assumption that only classes in the classpath are serialized/deserialized, all of their instance fields are non-nulls and can be allocated with any type that is safe. By applying these assumptions and relying on API modeling, our approach adds synthetic nodes into a previously computed call graph to improve its soundness with respect to serialization-related features.

We evaluated our approach with respect to its *soundness* (RQ1), *precision* (RQ2), *performance* (RQ3), and *usefulness* for a downstream client analysis (RQ4). We used 9 programs from the CATS Test Suite [Reif et al. 2018] and 10 projects from the XCorpus dataset [Dietrich et al. 2017b]. We compared our approach soundness and precision against off-the-shelf construction algorithms available on Soot [Vallée-Rai et al. 1999], Wala [WALA 2024], OPAL [Eichberg and Hermann 2014] and Doop [Bravenboer and Smaragdakis 2009].

In our experiments, we found that only the call graphs that used CHA or RTA could (partially) infer the callback methods that could arise at runtime. Our approach, on the other hand, provided support for all the callback methods in the serialization and deserialization . In an analysis by comparing runtime call graphs with the statically build call graphs, our approach introduced less spurious edges. Finally, by measuring the running times of our approach, compared with its counterpart call graph construction algorithm (Salsa and Wala), we found that our approach did not incur significant overhead.

## DATA-AVAILABILITY STATEMENT

The scripts and data to obtain the experimental results described in Section 4 and Seneca's implementation are available in our GitHub repository (https://github.com/s2e-lab/seneca/) and Zenodo (https://zenodo.org/doi/10.5281/zenodo.10464129).

## ACKNOWLEDGMENTS

This material is based upon work supported by the National Science Foundation under Grant No. CCF-1943300.